\documentclass[floatfix,nofootinbib,twocolumn,preprintnumbers,notitlepage,nolongbibliography]{revtex4-1}
\usepackage{times}
\usepackage{xcolor}
\usepackage{amsfonts}
\usepackage{amsmath}
\usepackage{amssymb}
\usepackage{bm}
\usepackage{dcolumn}
\usepackage{booktabs}
\usepackage{multirow}
\usepackage{rotating}
\usepackage{adjustbox}
\usepackage{array}

\usepackage{enumerate}
\usepackage{epsfig}
\usepackage{graphicx}
\usepackage{graphics}
\usepackage[latin1]{inputenc}
\usepackage{latexsym}
\usepackage{rotating}
\usepackage{hyperref}
\usepackage[toc,page]{appendix}
\definecolor{mgreen}{rgb}{0.1,0.7,0.1}

\DeclareMathOperator{\sign}{sign}

\begin{document}

\title{Effect of Type II Strong Gravitational Lensing on Tests of General Relativity}
\author{Purnima Narayan}
\affiliation{Department of Physics and Astronomy, The University of Mississippi, University, Mississippi 38677, USA}
\author{Nathan~K.~Johnson-McDaniel}
\affiliation{Department of Physics and Astronomy, The University of Mississippi, University, Mississippi 38677, USA}
\author{Anuradha Gupta}
\affiliation{Department of Physics and Astronomy, The University of Mississippi, University, Mississippi 38677, USA}

\date{\today}

\begin{abstract}

Gravitational wave (GW) observations of binary black hole (BBH) coalescences provide a unique opportunity to test general relativity (GR) in the strong-field regime. To ensure the reliability of these tests, it is essential to identify and address potential sources of error, particularly those arising from missing physics in the waveform models used in GW data analysis. This paper investigates potential biases in these tests arising from strong gravitational lensing, an effect not currently incorporated into the standard framework for GR tests. In the geometric optics approximation, strong lensing produces three types of images: Type~I, Type~II, and Type~III. While Type~I and Type~III images do not distort the signal, Type~II images introduce a characteristic phase shift that can mimic GR deviations for signals with higher-order modes, precession, or eccentricity. We assess the response of four standard GR tests on simulated Type~II lensed BBH signals, including the two parameterized tests (TIGER and FTI), the modified dispersion relation test and the inspiral-merger-ringdown consistency test. We focus on precessing waveforms for binaries with total masses of $20M_{\odot}$ and $80M_{\odot}$, and dimensionless spins of $0.5$ and $0.95$, considering a fixed signal-to-noise ratio of $25$ using the design A$+$ sensitivity of the LIGO-Virgo network. Our findings indicate that more mass-asymmetric and higher-spin binaries show larger false deviations from GR in the TIGER and modified dispersion relation tests when applying GR tests to Type~II lensed signals. These results highlight the risk of false GR violations as detector sensitivity improves in future observational runs. Therefore, it is crucial to consider the possibility of strong lensing before drawing conclusions about deviations from GR in GW signals.

\end{abstract}

\maketitle

%%%%%%%%%%%%%%%%%
\section{Introduction}
\label{sec:intro}
%%%%%%%%%%%%%%%%%

Einstein's theory of general relativity (GR) has been extensively tested and validated in weak-field, low-speed, and linear gravity regimes~\cite{Will:2014kxa,Wex:2014nva,Will:2014kxa,Wex:2014nva,Weisberg:2004hi,Voisin:2020lqi,Kramer:2021jcw}. However, the detection of gravitational waves (GWs) from binary black hole (BBH) mergers provides a unique opportunity to probe GR in the highly nonlinear, strong-field regime, an otherwise inaccessible domain. This regime has been the focus of several tests over the years, none of which have revealed statistically significant deviations from GR~\cite{GW150914_TGR,GW170817_TGR,O2_TGR,O3a_TGR,O3b_TGR}. Testing GR using GWs relies on waveform models that are compared with the detected GW data. The current tests of GR carried out by the LIGO-Virgo-KAGRA (LVK) collaboration, are based on waveform models that neglect certain physical effects that could be present in the signal and give rise to false deviations from GR~\cite{Gupta:2024gun}. One such effect is strong gravitational lensing, wherein a mass located between the source and observer bends the GW signal, potentially producing multiple images with different magnifications and arrival times (see, e.g.,~\cite{Schneider:1992bmb,Narayan:1996ba,Dai:2016igl,Ng:2017yiu,Oguri:2018muv}). Another effect that is not currently included in GW tests of GR carried out by the LVK is eccentricity, and these tests are indeed found to produce spurious deviations from GR when applied to signals with nonnegligible eccentricity~\cite{Saini:2022igm, Bhat:2022amc, Narayan:2023vhm, Saini:2023rto, Shaikh:2024wyn}.

We focus on the effect due to Type~II lensing that causes nontrivial deformations of the signal~\cite{Dai:2017huk, Ezquiaga:2020gdt}, as discussed below. Several techniques have been developed to identify strongly lensed GW signals (see, e.g.,~\cite{Haris:2018vmn,Li:2019osa,2023PhRvD.107l3015L,Janquart:2021qov,Goyal:2021hxv,Janquart:2023osz,Ezquiaga:2023xfe,Li:2023zdl,Magare:2024wje,Chakraborty:2024net,Offermans:2024fot,Barsode:2024zwv}). While searches for strong lensing in GW data have yet to detect a definitive signature~\cite{Hannuksela:2019kle,Dai:2020tpj,LIGOScientific:2021izm,LIGOScientific:2023bwz}, the likelihood of detecting such events will increase as detectors become more sensitive. For instance, for the LVK network with the LIGO detectors at A+ sensitivity and Virgo and KAGRA at design sensitivity, Ref.~\cite{Wierda:2021upe} finds that one could expect several strongly lensed events per year, while third-generation observatories such as Cosmic Explorer (CE)~\cite{Reitze:2019iox} and the Einstein Telescope (ET)~\cite{Punturo:2010zz} could each detect over $100$ lensed events per year~\cite{Wang:2021kzt}. Thus, one has to consider the possibility of a Type~II lensed signal potentially leading to a false deviation from GR. While one expects the LVK lensing analyses (as in~\cite{LIGOScientific:2021izm,LIGOScientific:2023bwz}) to be able to detect the effects of Type~II lensing in many cases (since they specifically look for the dephasing present in Type~II signals), as well as to potentially be able to identify companion lensed images to a Type~II lensed signal, it is still important to check the response of standard tests of GR to such a signal, particularly since the tests of GR are carried out simultaneously with the lensing analyses.

One obtains Type~II signals in the geometric optics treatment of gravitational lensing, which has been extensively studied in the literature (see, e.g., \cite{Schneider:1992bmb}). In the short-wavelength, geometric optics regime, such as the lensing of GWs by galaxies, the Kirchhoff diffraction integral describing the effects of lensing simplifies to a Gaussian integral around extremal points on the image plane, corresponding to three types of images: Type~I (minimum), Type~II (saddle), and Type~III (maximum)~\cite{Ezquiaga:2020gdt}. Of these, Type~II images introduce a non-trivial distortion to the waveform in all but specialized cases, such as highly symmetric binaries~\cite{Ezquiaga:2020gdt}. Given that the probability of having a Type~II image in a strongly lensed merger is greater than $99.99\%$~\cite{Wang:2021kzt}, it is crucial to investigate their effects in order to properly assess any potential deviations from GR.

Previous studies have highlighted biases in parameter estimation when Type~II lensed images are analyzed using unlensed waveform models~\cite{Vijaykumar:2022dlp}. Additionally, biases in tests of GR arising from neglecting microlensing~\cite{Mishra:2023vzo} and millilensing~\cite{Liu:2024xxn} effects have been investigated. Recent studies have also explored the connection between the phase shift introduced by Type~II lensing and GR deviations, showing that GR deviations can be flagged by pipelines designed to identify Type~II lensed images~\cite{Wright:2024mco,Ezquiaga:2022nak}. Our work extends these findings, demonstrating that the application of standard LVK tests of GR to Type~II lensed signals can result in false positives, indicating strong GR deviations. Therefore, before any claims of a GR violation are made, it is essential to rule out the possibility of the signal originating from a strongly lensed binary.

Specifically, we consider the Test Infrastructure for GEneral Relativity (TIGER)~\cite{TIGER2014,Meidam:2017dgf,Roy:2025gzv}, Flexible-Theory-Independent (FTI)~\cite{Mehta:2022pcn}, Modified Dispersion Relation (MDR)~\cite{Mirshekari:2011yq,Baka:2025drk}, and Inspiral-Merger-Ringdown (IMR) consistency~\cite{Ghosh:2016qgn,Ghosh:2017gfp} tests and check their response to simulated Type II lensed BBH GW signals in the LIGO-Virgo network at its design O5 (A+) sensitivity~\cite{Aasi:2013wya}. We consider signals from precessing BBHs with redshifted masses of $20M_{\odot}$ and $80M_{\odot}$, each with mass ratios of $2,5$ and spins of $0.5, 0.95$, and all having an inclination angle of $\pi/3$. Each binary is placed at a luminosity distance (with unit magnification) so that it has a network signal-to-noise ratio (SNR) of 25.
For comparison, we also apply these tests to the unlensed versions of these signals for selected testing parameters.

We find TIGER to be the most sensitive to Type II lensing effects. The absence of precession in current FTI implementations results in notable systematics for both lensed and unlensed signals. For each binary configuration, the testing parameter corresponding to frequency independent phase shift shows the strongest support for MDR. And finally for the IMR consistency test, we focus on $80M_{\odot}$ binaries due to SNR considerations, finding that while the lower mass ratio binaries remain consistent with GR, the higher mass ratio binaries show GR inconsistencies in both lensed and unlensed cases. We also investigate the nature of the GR parameters when analyzing lensed waveforms with unlensed waveform models and find biases in the source parameters for the TIGER, FTI, and MDR tests.

This paper is organized as follows: In Sec.~\ref{sec:theory}, we explain the effect of the strong lensing on GWs and describe the method followed to simulate Type~II lensed signals. In Sec.~\ref{sec:inj}, we give the specifics of our simulated observations and in Sec.~\ref{sec:tgr}, we give the details of the four tests of GR we consider. In Sec.~\ref{sec:results}, we discuss the results from our analysis and we conclude in Sec.~\ref{sec:concl}. We use geometrized ($G = c = 1$) units throughout.

%%%%%%%%%%%%%%%%%
\section{Theory}
\label{sec:theory}
%%%%%%%%%%%%%%%%%

In this section we discuss how Type II lensed GW signals are simulated for our study and how the presence of higher modes or precession makes the Type~II phase shift lead to a nontrivial distortion of the waveform.

In the frequency domain and geometric optics approximation, the strongly lensed GW signal is~\cite{Ezquiaga:2020gdt}
\begin{equation}
 \begin{split}
	 \tilde{h}^{L}_{+,\times}(f) = \sum_j \left|\mu_j\right|^{1/2} \exp \left( 2\pi i f t_d - in_{j}\frac{\pi}{2}\sign(f) \right) \tilde{h}_{+,\times}(f),
 \end{split}
 \end{equation}
where $+$ and $\times$ represent the plus and cross polarizations; $\mu_{j}$ is the magnification of the $j^\text{th}$ image; \( f \) is the frequency of the GW; \( t_d \) is the geometric time delay caused by lensing; the Morse index $n_{j}$ takes the values $0$, $1$, and $2$ for Type I, II, and III images respectively; and $\tilde{h}_{+,\times}(f)$ is the unlensed frequency-domain GW signal. 

For a Type~I image, there is no extra phase shift, and the lensed waveform is equivalent to the unlensed one apart from the magnification and $2\pi f t_d$ term, which accounts for the travel time to the observer and cannot be measured in the GW signal.
For a Type~III image, the phase shift equals $\pi$ for all frequencies, resulting in a waveform that flips its sign. This sign flip can be compensated exactly by a $\pi/2$ shift in the polarization angle, as explained in Appendix~D of~\cite{Ezquiaga:2020gdt}.
A nontrivial effect occurs for a Type~II image, where all positive-frequency components are shifted by a phase $-\pi/2$, and all negative-frequency components are shifted by $\pi/2$. This results in an overall phase shift of $-(\pi/2)\sign(f)$, making Type~II images equivalent to a Hilbert transform of Type~I images. Since the GW strain is real-valued, we have $\tilde{h}(-f) = \tilde{h}^*(f)$, where the star denotes the complex conjugate. The Type~II phase shift maintains this relation, as it must, and we can thus restrict our attention to the positive frequencies, as is standard in GW data analysis, since the negative frequencies do not provide any additional information.
The Type~II lensed signal that we use to construct our lensed waveform then is
\begin{equation}\label{lensed_signal}
	\tilde{h}^{L, \text{II}}_{+,\times}(f) = -i\tilde{h}_{+,\times}(f),
\end{equation}
where we have taken the magnification to be unity. We make this assumption following studies that indicate that the distribution of magnifications of Type~II lensed images peaks close to 1~\cite{Wang:2021kzt}. Additionally, since the SNR and redshifted masses are fixed in our analysis, and we are considering BBHs, variations in magnification merely correspond to changes in the binary's luminosity distance.

We now consider the situations in which Type~II lensing leads to a nontrivial deformation of the signal (as discussed in~\cite{Ezquiaga:2020gdt}). We start by considering a quasicircular, aligned-spin system, where the frequency domain Type~II lensed signal with unit magnification is given (for $f>0$) by:
\begin{equation}
        \tilde{h}^{L,\text{II}, \parallel}_{+,\times}(f) = \sum_{\ell \geq 2} \sum_{m \geq 0}\mathcal{A}_{\ell m}(f) \mathcal{Y}_{\ell m,+,\times}(\iota)\exp\left[i\left(m\phi_c - \frac{\pi}{2}\right)\right],
\end{equation}
where $\parallel$ denotes that this just holds for aligned-spin systems, $\mathcal{A}_{\ell m}(f)$ represents the amplitude of each (spin-weighted spherical harmonic) mode, \(\phi_c\) is the coalescence phase, and $\mathcal{Y}_{\ell m,+,\times}(\iota)$ encodes the dependence on the inclination angle $\iota$. 

For a simple quasicircular, aligned-spin binary dominated by the $\ell = 2, m = 2$ mode, the $\pi/2$ phase shift from Type~II lensing can be absorbed into the coalescence phase $\phi_c$ via the shift $\phi_c \rightarrow \phi_c + \pi/4$. However, the phase shift required to absorb the lensing effect depends on the mode number $m$, so if there is a nonnegligible contribution from a mode with $m \neq 2$, the $\phi_c$ shift for the $m = 2$ mode will not provide the shift of $\phi_c \rightarrow \phi_c + \pi/(2m)$ necessary to account for the lensing phase shift seen in the $m \neq 2$ mode. 

This also holds if a signal from a precessing binary contains only the $(2,2)$ mode in the co-precessing frame (this is never the case physically, but considering this situation provides useful intuition). In such a case, Type~II lensing can still lead to detectable effects in the inertial frame. Specifically, the binary's precessional motion causes the $\ell = 2$ modes with $m \neq 2$ to be nonzero in the inertial frame. The contributions of different $m$ modes to the waveform have distinct dependencies on the azimuthal angle $\phi$ [cf.\ Eqs.~(2.1) and~(2.2) in~\cite{Pratten:2020fqn}]. As a result, a shift in $\phi$ that compensates for the Type~II lensing phase shift in the $|m| = 2$ modes will not produce the correct phase shift in modes with $|m| \neq 2$, making the lensing-induced phase shift observable. Intuitively, this arises because for a non-precessing binary, changing $\phi$ is equivalent to adjusting the phase at coalescence, since the orbital plane is fixed in space. In contrast, for a precessing binary, a change in the coalescence phase corresponds to a more complex, time-dependent change in $\phi$ at different points in the evolution, due to the precession of the orbital plane.

The mode-dependent phase shift causes a mismatch in the phase evolution across different modes, resulting in a nontrivial distortion of the overall waveform, illustrated in the time domain in, e.g.,~\cite{Wang:2021kzt,Vijaykumar:2022dlp}. A binary exhibiting observable precession will therefore have modes with different $|m|$ values present in the inertial frame and will experience such nontrivial distortions due to Type~II lensing. We thus focus in this study on precessing, mass-asymmetric binaries that have significant higher-mode content. Eccentric waveforms also exhibit nontrivial distortions from Type~II lensing, as discussed in~\cite{Ezquiaga:2020gdt}, but we do not consider them here.

%%%%%%%%%%%%%%%%%
\section{Simulated observations and parameter estimation setup}
\label{sec:inj}
%%%%%%%%%%%%%%%%%

We simulate Type~II lensed BBH image strains in the LIGO-Virgo network by applying Eq.~(\ref{lensed_signal}) to standard waveform models (discussed below) for unlensed signals from quasicircular, precessing BBHs. We consider the network at the sensitivity predicted for O5/A+~\cite{Aasi:2013wya} (with the more sensitive Virgo noise curve) when analyzing the observations. We do not include KAGRA~\cite{KAGRA:2018plz} owing to the significant uncertainties in its plus-era sensitivity (see~\cite{timeline_graphic}). To avoid biases from particular noise realizations, we do not incorporate noise in our simulated observations (i.e., we take the zero realization of Gaussian noise). We will also use the standard GW data analysis terminology of referring to these simulated GW observations as injections.

We select two (redshifted) total masses, 20\(M_{\odot}\) and 80\(M_{\odot}\), to examine the lensing effects on an inspiral-dominant binary and one where higher-order modes have a greater impact on the inference. For the lower mass (20\(M_{\odot}\)), the signal falls in the detectors' most sensitive range mainly during the inspiral phase with at most 6 precessional cycles in band. In contrast, for the 80\(M_{\odot}\) binary, the merger and ringdown phases lie in a more sensitive frequency range for the detectors with just 1 precessional cycle in band. Since the amplitudes of higher modes are larger in the merger-ringdown phase than earlier in the inspiral, they thus have a larger effect on the parameter estimation (PE) in this case, though they remain subdominant.

For each total mass, we study two mass ratios $q \in \{2, 5\}$ and two dimensionless spin magnitudes such that $\chi_1=\chi_2=\chi \in \{0.5, 0.95\}$. In total, we consider simulated observations of $8$ Type~II lensed binaries, each with an inclination angle of $\pi/3$ and spin tilt angles of ($2.03$, $0.43$)~rad, where all spin directions are specified at $20$~Hz. We choose the component of the spin of the primary in the orbital plane to be in the direction of the vector from the smaller to the larger black hole, and the difference between the azimuthal angles of the individual spin vectors to be $\phi_{12} = 2.59$~rad. For the extrinsic parameters, for each binary we choose the luminosity distance (with unit magnification) such that the network SNR is $25$ for the fixed randomly chosen values for the other extrinsic parameters we use: right ascension ($5.41$~rad), declination ($0.88$~rad), polarization angle ($2.59$~rad), coalescence GPS time ($1126259642.413$~s), and coalescence phase ($2.27$~rad). We give the mass ratios, spins and luminosity distances of the simulations considered in Table~\ref{tab:sims}.

\begin{table}
	\caption{\label{tab:sims}The lensed simulations we consider and their properties. All the binaries have a luminosity distance chosen such that the network SNR is $25$. The dimensionless spins of the two black holes are the same.} 
\begin{tabular}{*{6}{c}}
\hline\hline
	$q$ & & & $\chi$ & & $D_{L}$ (Gpc)\\
\hline
	\multicolumn{6}{c}{$M = 20M_{\odot}$}\\
\hline
	$2$ & & & $0.5$ & & $0.58$\\
	$2$ & & & $0.95$ & & $0.62$\\
	$5$ & & &$0.5$ & & $0.54$\\
	$5$ & & & $0.95$ & & $0.43$\\
\hline
	\multicolumn{6}{c}{$M = 80M_{\odot}$}\\
\hline
	$2$ & & & $0.5$ & & $1.95$\\
	$2$ & & & $0.95$ & & $2.15$\\
	$5$ & & & $0.5$ & & $1.57$\\
	$5$ & & & $0.95$ & & $0.83$\\
\hline\hline
\end{tabular}
\end{table}
%\begin{table}
%	\caption{\label{tab:snrs}A summary of the SNRs from the co-precessing modes of the waveform models considered in this study. We present the $2,\pm2$, $(2,\pm1)$, $(3,\pm3)$, $(3,\pm2)$, and $(4,\pm4)$ modes for \texttt{IMRPhenomXPHM} and the $2,\pm2$, $(2,\pm1)$, $(3,\pm3)$, and $(4,\pm4)$ modes for \texttt{SEOBNRv4PHM}. We do not consider the $(5,\pm5)$ mode for \texttt{SEOBNRv4PHM} since they do not make a significant contribution to the signal.} 
%\begin{tabular}{*{6}{c}}
%\hline\hline
%	$q$ & & & $\chi$ & & $D_{L}$ (Gpc) & \texttt{SEOBNRv4HM\_ROM} (22,21, 33, 44) & \texttt{IMRPhenomXPHM} (22,21,33,32,44)\\
%\hline
%	\multicolumn{6}{c}{$M = 20M_{\odot}$}\\
%\hline
%	$2$ & & & $0.5$ & & $0.58$ & $25.6, 1.4, 2.6, 0.8$ & $24.9,1.4,2.5,0.2,0.8$\\
%	$2$ & & & $0.95$ & & $0.62$ & $25.6, 2.2, 2.3, 0.6$ & $24.9,1.7,0.2,0.2,0.7$\\
%	$5$ & & &$0.5$ & & $0.54$ & $24.7, 2.3, 3.6, 0.8$ & $24.5,1.4,0.3,0.4,0.8$\\
%	$5$ & & & $0.95$ & & $0.43$ & $24.1, 3.2, 4.2, 1.1$ & $24.5,2.4,0.4,0.4,1.2$\\
%\hline
%	\multicolumn{6}{c}{$M = 80M_{\odot}$}\\
%\hline
%	$2$ & & & $0.5$ & & $1.95$ & $22.4, 1,7, 3.2, 1.2$ & $24.2,1.6,3.5,0.2,1.3$\\
%	$2$ & & & $0.95$ & & $2.15$ & $22.9, 2.9, 2.6, 1.1$ & $24.4,2.5,3.0,0.2,1.1$\\
%	$5$ & & & $0.5$ & & $1.57$ & $23.6, 5.7, 3.9, 1.1$ & $23.9,1.5,4.5,0.9,1.2$\\
%	$5$ & & & $0.95$ & & $0.83$ & $-,-,-,-$ & $23.2,6.1,7.3,1.7,2.8$\\
%\hline\hline
%\end{tabular}
%\end{table}

\begin{table}[htbp]
	\caption{\label{tab:snr} SNRs for co-precessing modes of the waveform models \texttt{SEOBNRv4PHM} (\texttt{SEOB}) and \texttt{IMRPhenomXPHM} (\texttt{IMRPh}) for the simulated observations used in this study. Each row corresponds to a BBH system characterized by total mass \(M\), mass ratio \(q\), and spin magnitude \(\chi\). We quote SNRs for the $(2,\pm2)$, $(2,\pm1)$, $(3,\pm3)$, and $(4,\pm4)$ modes that are present in both waveform models and also the $(3,\pm2)$ modes that are included in \texttt{IMRPhenomXPHM}. We do not consider the $(5,\pm5)$ modes for \texttt{SEOBNRv4PHM} in our study since they do not make a significant contribution to the signal. We omit the $\pm$ in the mode labels in the table for brevity.}
\begin{adjustbox}{max width=0.48\textwidth}
{\begin{tabular}{@{}llcccccc@{}}
%\toprule
	\hline\hline \\
	\multicolumn{2}{c}{\textbf{System Parameters}} & \multicolumn{1}{c}{\textbf{Model}} & \multicolumn{5}{c}{\textbf{Mode SNRs}} \\
\cmidrule(lr){1-2} \cmidrule(lr){3-3} \cmidrule(lr){4-8}\\
	$M~(M_\odot)$ & \hspace{0.1cm}$q$, $\chi$ & & $(2,2)$ & $(2,1)$ & $(3,3)$ & $(3,2)$ & $(4,4)$ \\
\\
\multirow{4}{*}{     20}
	& \multirow{2}{*}{$2$, $0.5$} & \texttt{SEOB}   & 25.6 & 1.4 & 2.6 & --  & 0.8 \\
	&                                  & \texttt{IMRPh}  & 24.9 & 1.4 & 2.5 & 0.2 & 0.8 \\
\cline{2-8}
	& \multirow{2}{*}{$2$, $0.95$} & \texttt{SEOB}   & 25.6 & 2.2 & 2.3 & --  & 0.6 \\
	&                                   & \texttt{IMRPh}  & 24.9 & 1.7 & 2.3 & 0.2 & 0.6 \\
\hline

\multirow{4}{*}{    20}
	& \multirow{2}{*}{$5$, $0.5$} & \texttt{SEOB}   & 24.7 & 2.3 & 3.6 & --  & 0.8 \\
	&                                   & \texttt{IMRPh}  & 24.5 & 1.4 & 3.6 & 0.6 & 1.0 \\
\cline{2-8}
	& \multirow{2}{*}{$5$, $0.95$} & \texttt{SEOB}   & 24.1 & 3.2 & 4.2 & --  & 1.1 \\
	&                                    & \texttt{IMRPh}  & 24.5 & 2.4 & 4.5 & 0.3 & 1.2 \\
\hline

\multirow{4}{*}{    80}
	& \multirow{2}{*}{$2$, $0.5$} & \texttt{SEOB}   & 22.4 & 1.7 & 3.2 & --  & 1.2 \\
	&                                   & \texttt{IMRPh}  & 24.2 & 1.6 & 3.4 & 0.5 & 1.3 \\
\cline{2-8}
	& \multirow{2}{*}{$2$, $0.95$} & \texttt{SEOB}   & 22.9 & 2.9 & 2.6 & --  & 1.1 \\
	&                                    & \texttt{IMRPh}  & 24.3 & 2.6 & 3.0 & 0.4 & 1.1 \\
\hline

\multirow{4}{*}{    80}
	& \multirow{2}{*}{$5$, $0.5$} & \texttt{SEOB}   & 23.6 & 5.7 & 3.9 & --  & 1.1 \\
	&                                   & \texttt{IMRPh}  & 23.9 & 1.5 & 4.4 & 1.0 & 1.2 \\
\cline{2-8}
	& \multirow{2}{*}{$5$, $0.95$} & \texttt{SEOB}   & 29.2   & 8.1  & 8.7  & --  & 3.4  \\
	&                                    & \texttt{IMRPh}  & 23.2 & 6.1 & 7.3 & 1.7 & 2.8 \\
\hline\hline
\end{tabular}}
\end{adjustbox}
\end{table}

We analyze all injections starting from a lower Fourier frequency of $20$~Hz, with upper limits of $896$~Hz for the $20M_{\odot}$ cases and $448$~Hz for the $80M_{\odot}$ cases, except where different frequency limits are applied in the inspiral and postinspiral analyses for the IMR consistency test. The upper frequencies are chosen to be sufficiently high that the SNR in the frequencies above the cutoff is negligible. Specifically, they come from the Nyquist frequencies associated with the sampling frequencies of $2048$~Hz and $1024$~Hz used for the $20M_{\odot}$ and $80M_{\odot}$ cases, respectively, after accounting for the roll-off factor of $0.875$ that accounts for the effects of a window function (as discussed in Appendix~E of~\cite{KAGRA:2021vkt}).
Parameter estimation is performed using \texttt{\textsc{Bilby}}~\cite{Ashton:2018jfp} with the nested sampler \texttt{\textsc{Dynesty}}~\cite{Speagle_2020}, and the \texttt{\textsc{bilby\_tgr}}~\cite{2024zndo..10940210A} plugin for the TIGER, FTI, and MDR tests.
We use uniform priors for the component masses, spin magnitudes and non-GR parameters; isotropic priors for spin directions, binary orientation, and sky location; and a luminosity distance prior corresponding to a uniform merger rate in the source's comoving frame, using the same Planck cosmology~\cite{Planck:2015fie} as in the LVK catalog analysis~\cite{KAGRA:2021vkt}.

%%%%%%%%%%%%%%%%%
\section{Tests of GR}
\label{sec:tgr}
%%%%%%%%%%%%%%%%%

The tests we conduct are based on waveform models for unlensed GW signals from quasicircular BBHs in GR. For these tests, we use the \texttt{IMRPhenomXPHM}~\cite{Pratten:2020ceb} model for both injection and recovery in the TIGER, MDR, and IMR consistency tests, while in the FTI test, we use the \texttt{SEOBNRv4PHM}~\cite{Ossokine:2020kjp} model for injection and \texttt{SEOBNRv4HM\_ROM}~\cite{Cotesta:2018fcv,Cotesta:2020qhw} for recovery. We use the aligned-spin model \texttt{SEOBNRv4HM\_ROM} since it forms the basis of the only higher-mode version of the FTI test currently implemented, and thus use the closest precessing waveform model for the injections, \texttt{SEOBNRv4PHM}. We give the SNRs for the co-precessing modes of the simulated observations considered in this study for the waveform models \texttt{IMRPhenomXPHM} and \texttt{SEOBNRv4PHM} in Table~\ref{tab:snr}.

\texttt{IMRPhenomXPHM} is a frequency-domain phenomenological model for quasicircular BBH waveforms that enhances the accuracy of the baseline \texttt{IMRPhenomXAS} model~\cite{Pratten:2020fqn} for the dominant $(2,\pm2)$ modes of aligned-spin waveforms by incorporating spin precession and the subdominant modes $(2,\pm1)$, $(3,\pm3)$, $(3,\pm2)$, and $(4,\pm4)$ in the coprecessing frame. \texttt{SEOBNRv4HM\_ROM} is a frequency-domain reduced-order model for quasicircular BBH waveforms based on a time-domain aligned-spin effective-one-body (EOB) model, including the $(2,\pm1)$, $(3,\pm3)$, $(4,\pm4)$, and $(5,\pm5)$ subdominant modes. However, we do not consider the $(5, \pm 5)$ modes in our FTI analysis, since they do not make a significant contribution to the signal. To evaluate the impact of excluding the $(5, \pm 5)$ modes, we computed the SNR difference using the \texttt{SEOBNRv4PHM} waveform model and found the difference to be less than $0.2$ and $0.3$ for the $M=20M_\odot$ and  $M=80M_\odot$ binaries respectively. This difference is small enough to not impact our results significantly.

%--------------
\subsection{TIGER and FTI}
\label{ssec:par}
%--------------
Parameterized tests of GR have been developed to detect potential deviations from GR by introducing modifications to the frequency-domain phase coefficients of GR waveform models. One such framework is the TIGER test, which introduces deviations in the GR phase coefficients of \texttt{IMRPhenomXAS}, a non-precessing dominant-mode BBH waveform model. This model is then extended to higher modes~\cite{Garcia-Quiros:2020qpx} following which the aligned-spin higher-mode model is twisted up to account for precession, giving the final modified \texttt{IMRPhenomXPHM} model used in our study. In the inspiral portion of the signal, the deviations introduced in the post-Newtonian (PN) coefficients also affect the higher modes (in the coprecessing frame), as illustrated below. However, deviations in the phenomenological intermediate and merger-ringdown coefficients only affect the dominant mode of the waveform (in the coprecessing frame). The end of the inspiral portion of the signal is set around the frequency of the minimum energy circular orbit and is different for different modes~\cite{Garcia-Quiros:2020qpx}.

The deviations to the GR phase are introduced into the PN coefficients (\(\varphi_k\) and \(\varphi_{k(l)}\)) within the Fourier-domain inspiral phase of the waveform. The inspiral phase \(\Phi_{\ell m}(f)\) of the $(\ell, m)$ spin-$(-2)$-weighted spherical harmonic mode in the coprecessing frame is given by
\begin{equation}
    \label{waveformPhase}
    \begin{split}
        \Phi_{\ell m}(f) &= \frac{3}{128\eta v^5}\frac{|m|}{2}  \sum_{n=0}^7\left( \varphi_k + 3\varphi_{k(l)} \log v \right) v^k
    \end{split}
\end{equation}
(excluding additive constants and phase and time shifts), where the factor of $3$ in the logarithmic term arises from the definition of PN coefficients used in TIGER. Here, \(\eta = m_1 m_2/M^2\) is the symmetric mass ratio, with \(m_1\) and \(m_2\) representing the binary's individual (redshifted) masses, $M = m_1 + m_2$ is the total (redshifted) mass of the binary. The variable \(v\) is defined as \(v = (2\pi Mf/|m|)^{1/3}\), where \(f\) is the GW frequency. The logarithmic coefficients \(\varphi_{k(l)}\) are non-zero only for \(k \in \{5, 6\}\). We also consider $-1$PN deviations that are not present in GR, but would appear in the above expression with $k=-2$.

As discussed in~\cite{Roy:2025gzv}, the TIGER deviations are introduced in both the PN coefficients of the inspiral phase and in the phenomenological coefficients $b_k$ and $c_k$ during the intermediate and merger-ringdown phases of the waveform, respectively. The intermediate parameters $b_k$ ($k\in\{1,2,3,4\}$) have a frequency dependence $f^{1-k}$ ($\log f$ for $k = 1$) and the merger-ringdown parameters $c_1, c_2, c_4$ have the dependence $f^{2/3}, f^{-1}, f^{-3}$ respectively and $c_l$ is the coefficient of the antiderivative of the Lorentzian that models the ringdown. In the \texttt{IMRPhenomXAS} model, the phase is constructed to be $C^1$ continuous, such that any modification in the lower-frequency (e.g., inspiral) phase also affects the higher-frequency portions of the phase (the intermediate and merger-ringdown regions for inspiral deviations). All the deviations are parameterized by introducing a deviation parameter \(\delta \hat{p}_k\) through the substitution \(p_k \to (1 + \delta \hat{p}_k)p_k\), where \(p_k\) denotes any of the PN and phenomenological coefficients. An exception is made for $\delta \hat{\varphi}_{-2}$ and $\delta \hat{\varphi}_1$, which are zero in GR, and thus are normalized by just the $0$PN coefficient. The normalization of the PN coefficients uses the coefficient without the spin contributions to avoid degeneracies since the coefficients can become zero when including the spin contributions. In GR, all deviation parameters are zero by definition.

In the standard TIGER test, only one PN parameter is varied at a time to avoid generating uninformative results, as is illustrated for GW150914 in~\cite{GW150914_TGR}. However, deviations affecting multiple PN coefficients can still be detected by individually varying the parameters, even if the specific coefficients are not directly modified, as is shown in~\cite{Meidam:2017dgf,TGR_relation}. Nevertheless, with multiband observations of BBHs with space-based and ground-based GW detectors, the degeneracies between parameters will be removed and it will be possible to constrain all PN coefficients simultaneously~\cite{Gupta:2020lxa,Datta:2020vcj}. And even for observations in a single band, one can also obtain the well-measured combinations of parameters using principal component analysis~\cite{Shoom:2021mdj,Saleem:2021nsb,Datta:2022izc,Datta:2023muk}. 

The FTI test~\cite{Mehta:2022pcn} is similar to TIGER but focuses exclusively on deviations in PN coefficients and can be applied to any aligned-spin waveform model. However, the current implementation of the higher-mode version is limited to \texttt{SEOBNRv4HM\_ROM}. Additionally, FTI tapers the deviations to zero beyond a certain frequency instead of allowing them to affect the rest of the signal. The same taper frequency used in the FTI analysis~\cite{Sanger:2024axs} of GW230529 is being used here. Even though the current implementation for FTI does not account for spin-precession in the waveform model, it is applied to all BBHs in the LVK testing GR analyses~\cite{O2_TGR,O3a_TGR,O3b_TGR} that satisfy its selection criteria. Specifically, FTI is applied to BBHs that have an inspiral network SNR $\geq 6$.

%--------------
\subsection{Modified dispersion relation}
\label{ssec:mdr}
%--------------

The MDR test~\cite{Baka:2025drk} constrains dispersive GW propagation, where different wavelengths travel at different velocities, leading to frequency-dependent modifications to the phase of the observed GW waveform. Specifically, the MDR test introduces a phenomenological dispersion relation
$E^2 = p^2 + A_{\alpha} p^{\alpha},$
where \(E\) and \(p\) represent the energy and momentum of the GWs, respectively, while \(A_\alpha\) and \(\alpha\) are phenomenological parameters that determine the strength of the deviation from GR and the frequency dependence of the dispersion, respectively, following~\cite{Mirshekari:2011yq}. For \(\alpha = 0\) and \(A_0 > 0\), this corresponds to the dispersion relation of a massive graviton.

Following the discussion in~\cite{O2_TGR}, we assume that the waveform near the source remains very close to that predicted by GR, so that the only modifications we consider are those arising from the dispersive propagation. This dispersion relation leads to an additional term proportional to \(A_\alpha f^{\alpha - 1}\) in the waveform's frequency-domain phase; the explicit group velocity expression is given in \cite{Ezquiaga:2022nak}. As in the previous LVK analyses (e.g.,~\cite{O3b_TGR}), we consider \(\alpha \in \{0, 0.5, 1, 1.5, 2.5, 3, 3.5, 4\}\), omitting \(\alpha = 2\) since there is no dispersion in this case. In this study, we parameterize the phase corrections in terms of $A_{\text{eff}}$, defined (as in~\cite{Baka:2025drk}) as:
\begin{equation}
	A_{\text{eff}}=\frac{D_{\alpha}}{D_{L}}(1+z)^{\alpha-1}A_{\alpha},
\end{equation}
where, $D_{L}$ is the luminosity distance, $z$ is the redshift and $D_{\alpha}$ is a distance parameter, with an explicit expression given in~\cite{O2_TGR}.

%--------------
\subsection{IMR consistency test}
\label{ssec:imrct}
%--------------

The IMR consistency test~\cite{Ghosh:2016qgn,Ghosh:2017gfp} checks the consistency between the low- and high-frequency portions of the frequency-domain GW signal (roughly corresponding to the inspiral and postinspiral regimes) of a BBH. The division between these portions is made at a cut-off frequency $\textit f_c$ corresponding to the frequency of the innermost stable circular orbit (ISCO) of the final Kerr black hole, obtained from the GR analysis of the full signal~\cite{Bardeen:1972fi}. We follow the procedure used in the LVK analysis and calculate the cut-off frequency using the medians of the individual masses and spins. The test evaluates the consistency of the two portions by comparing the (redshifted) final mass $M_f$ and spin $\chi_f$ from each portion, yielding deviation parameters
\[
\frac{\Delta M_f}{\bar{M}_f} := 2\frac{M_f^\text{insp} - M_f^\text{postinsp}}{M_f^\text{insp} + M_f^\text{postinsp}}, \quad \frac{\Delta \chi_f}{\bar{\chi}_f} := 2\frac{\chi_f^\text{insp} - \chi_f^\text{postinsp}}{\chi_f^\text{insp} + \chi_f^\text{postinsp}},
\]
where the ``insp'' and ``postinsp'' superscripts correspond to the low- and high-frequency portions of the signal. These deviation parameters should both be zero if the signal is consistent with the waveform model used in the analysis, which is a quasi-circular BBH merger in GR for all current applications. As in~\cite{O3a_TGR, O3b_TGR}, we reweight to a flat prior in the deviation parameters to obtain the final results. We obtain the final mass and spin using the following procedure: We perform PE separately on each portion of the signal using a standard BBH waveform model (in this case, \texttt{IMRPhenomXPHM}), parameterized by the binary's initial masses and spins. We then compute the final mass and spin using an average of fits to numerical relativity (NR) simulations~\cite{Hofmann:2016yih,Healy:2016lce,Jimenez-Forteza:2016oae} (where the aligned-spin final spin fits are augmented by the contribution from in-plane spins~\cite{spinfit-T1600168}, though we do not evolve the spins, following the LVK applications of the test).

\begin{figure*}
\includegraphics[width=0.99\linewidth]{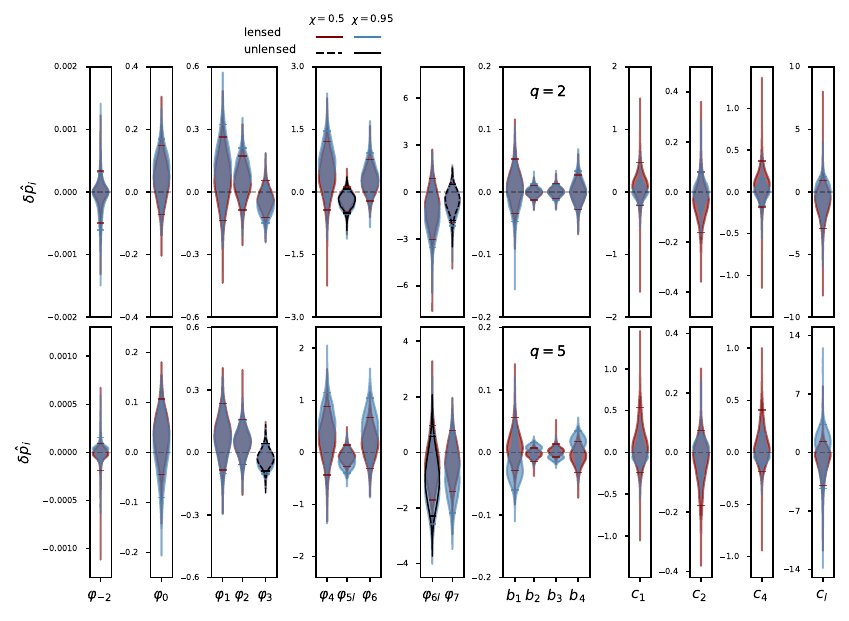}
	\caption{\label{fig:TIGER_plot_M20}The results of the TIGER test on the Type~II lensed simulated $M=20M_{\odot}$ injections, showing the results for mass ratios $q = 2$ and $5$ in the top and bottom panels, respectively, and the results for dimensionless spins $\chi=0.5$ and $0.95$ with different colours. The posteriors of the testing parameters are presented as violin plots and the associated $90\%$ credible intervals are shown using horizontal bars. We mark the GR value of zero with a dashed horizontal line. We also show the results for the unlensed injections we carried out for selected runs for each mass ratio and higher (lower) spin, as black solid (dashed) unfilled violins.}

\end{figure*}																					
\begin{figure*}
\includegraphics[width=0.99\linewidth]{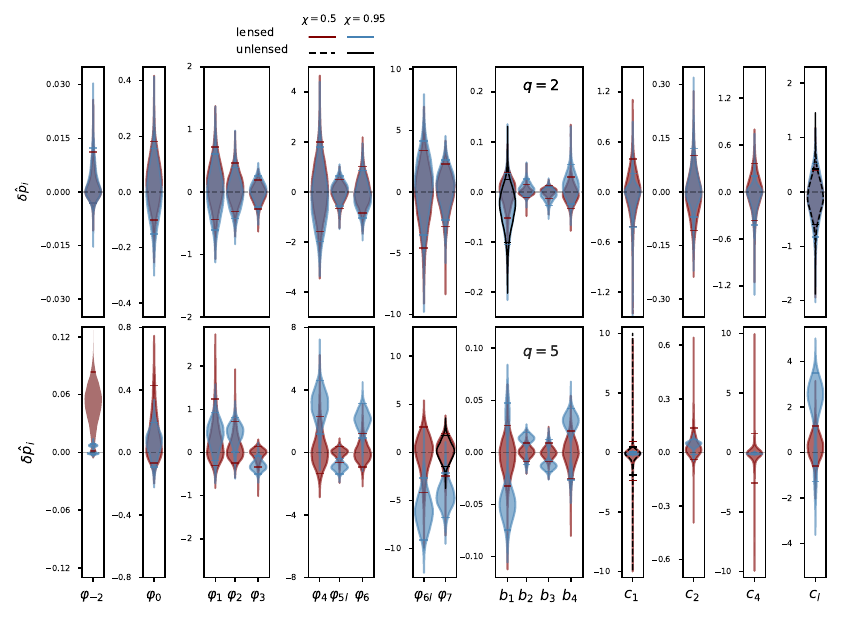}
	\caption{\label{fig:TIGER_plot_M80}The results of the TIGER test on the Type~II lensed simulated $M=80M_{\odot}$ injections. The design of the plot is the same as Fig.~\ref{fig:TIGER_plot_M20}.}

\end{figure*}																					
\begin{figure*}
\includegraphics[width=0.99\textwidth]{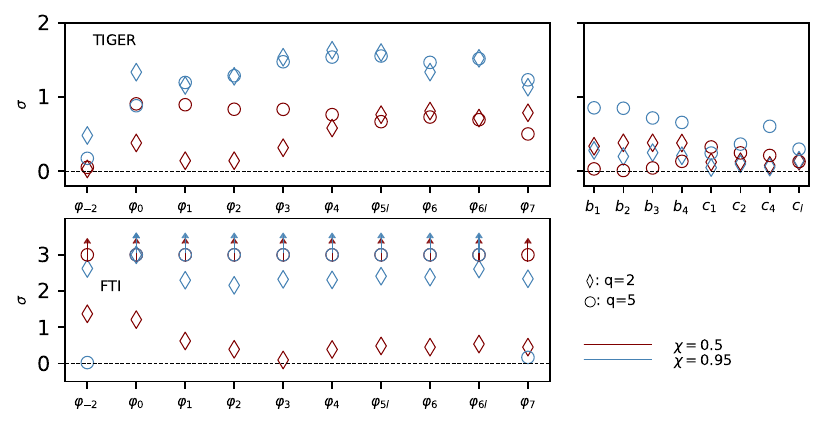}
	\caption{\label{fig:quantile_plot_m20} The Gaussian sigma value at which GR is excluded when applying the TIGER and FTI tests to the Type~II lensed injections with $M = 20M_{\odot}$. The upward arrows denote a lower bound of $3\sigma$, above which the values cannot be stated with certainty.}
\end{figure*}

\begin{figure*}
\includegraphics[width=0.99\textwidth]{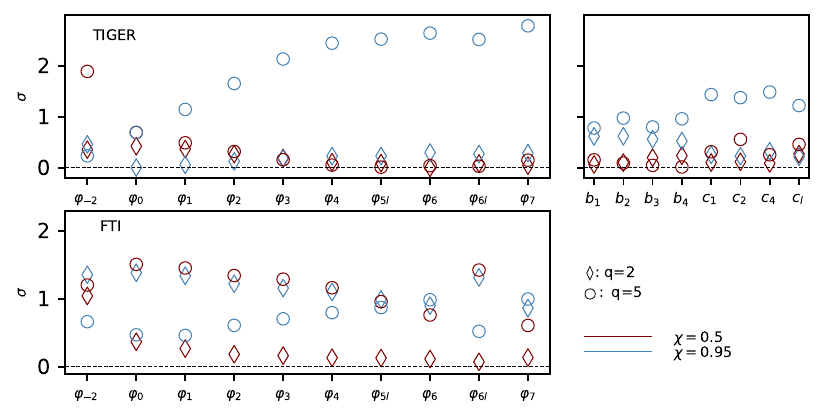}
	\caption{\label{fig:quantile_plot} The Gaussian sigma value at which GR is excluded when applying the TIGER and FTI tests to the Type~II lensed injections with $M = 80M_{\odot}$.}
\end{figure*}

%%%%%%%%%%%%%%%%%
\section{Results}
\label{sec:results}
%%%%%%%%%%%%%%%%%

We now discuss the results obtained when performing the four tests of GR described in Sec.~\ref{sec:tgr} on the simulated Type~II lensed signals discussed in Sec.~\ref{sec:inj}. Sometimes, results are expressed in terms of GR quantiles, which represent the quantile at which the GR value of the test is recovered. For the IMR consistency test, these quantiles are two-dimensional, with the GR quantile indicating the fraction of the posterior distribution enclosed by the isoprobability contour passing through the GR value. Larger GR quantile values correspond to larger deviations from GR. For all other tests, the GR quantiles are one-dimensional, where extreme values (either large or small) suggest that the posterior distribution does not peak near the GR value. 

\subsection{TIGER}
\label{ssec:tiger}

We give the posterior probability distributions (henceforth posterior distributions or posteriors) of the TIGER testing parameters for all our low and high mass injections in Figs.~\ref{fig:TIGER_plot_M20} and~\ref{fig:TIGER_plot_M80} respectively. We also summarize the statistical level at which GR is excluded in Figs.~\ref{fig:quantile_plot_m20} and~\ref{fig:quantile_plot}. We used the updated implementation of the TIGER pipeline, which fixes a bug that affected the $-1$PN results, where the deviation was not propagated to the higher modes, to analyze the $M = 80M_{\odot}$, $q = 5$ cases. This bug was only found after the completion of our initial analysis. Therefore, we applied the fix only to these two configurations that exhibited the most significant deviations from GR. The corrected analysis showed only a modest increase in the disfavoring of GR, with the overall structure of the posteriors remaining unchanged. Since these changes do not significantly affect the overall conclusions, we did not rerun with the updated implementation on the other injections considered in this study.

\subsubsection{$20M_{\odot}$ injections}\label{tiger_m20_results}

All lensed binaries with total mass $20 M_{\odot}$ are consistent with GR at $90\%$ credibility for all testing parameters. Additionally, we found that the posteriors of the GR parameters are almost identical to those obtained when analyzing these Type~II lensed signals with an unlensed GR waveform model, peaking at or close to the injected values. Additionally, as a check of these results, we consider one unlensed case for each injection. Here we select the testing parameter that yields the largest GR quantile. The testing parameters selected for the injections with $q=2$ are $\varphi_7$ and $\varphi_{5l}$, while for $q=5$ they are $\varphi_3$ and $\varphi_{6l}$; for both mass ratios the two cases correspond to $\chi=0.5$ and $0.95$, respectively. We plot the unlensed results in Fig.~\ref{fig:TIGER_plot_M20}, and notice that all their posteriors are in good agreement with the lensed ones. Thus, Type~II lensing does not lead to large GR deviations with TIGER in the lower mass cases we considered.

\subsubsection{$80M_{\odot}$ injections}
The TIGER analysis of the lensed binaries with total mass $80M_{\odot}$ we consider gives consistency with GR at the $90\%$ credible level in all cases except for the most extreme case of $q=5$ and $\chi=0.95$, where GR is even excluded at $>2\sigma$ for $\varphi_{3}$ through $\varphi_{7}$. 

We find that the cases where GR is excluded at $>2\sigma$ also give the largest differences in the recovery of the GR parameters, compared to the GR analysis. In particular, the posteriors peak at values larger than the injected ones for the chirp masses and luminosity distances, and at smaller values for $\theta_{1}$, placing them just outside the 90\% credible intervals. Here the chirp mass $(m_1m_2)^{3/5}/(m_1+m_2)^{1/5}$ is a combination of the component masses ($m_1$, $m_2$) which gives the leading effect of the masses on the GW signal, while $\theta_{1}$ is the the zenith angle between the Newtonian orbital angular momentum and the primary spin. In the GR analysis of the high-spinning binaries, the injected chirp mass falls within the $90\%$ credible interval. However, the injected values for luminosity distance, \(\theta_1\), and spin magnitudes lie just outside the $90\%$ credible interval.

As for the $20M_\odot$ injections, we apply the TIGER test on the unlensed injections by selecting the testing parameter that yields the largest GR quantile for each injection and compare their posteriors with the lensed counterparts.
The cases with the largest GR quantile selected for unlensed injections are \( c_{l} \), \( b_{1} \) for \( q=2 \), and \( c_{1} \), \( \varphi_{7} \) for \( q=5 \), where for both mass ratios the two cases correspond to $\chi=0.5$ and $0.95$, respectively. We plot the unlensed results in Fig.~\ref{fig:TIGER_plot_M80} and observe that the lensed and unlensed posteriors are almost identical in all cases except for the injection with \( q=5 \) and \( \chi=0.95 \), where the lensed case shows a GR deviation of $2.8\sigma$, while the unlensed posterior peaks at the GR value, as expected. Additionally, while we observe biases in the recovery of the GR parameters in the TIGER and GR analysis of the lensed cases, the corresponding unlensed injections recover the injected GR parameters around the median of the respective posteriors.

\subsection{FTI}

\begin{figure*}
\includegraphics[width=0.99\linewidth]{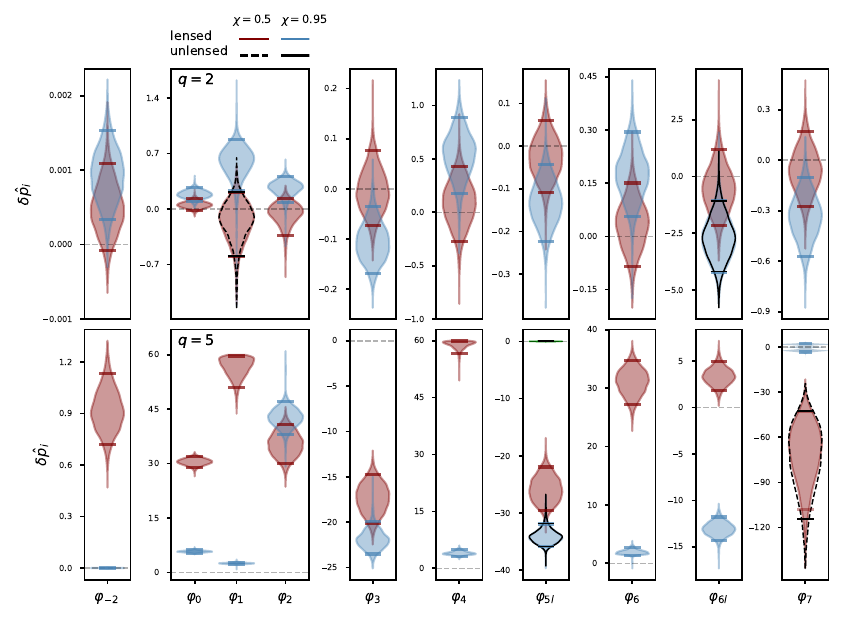}
	\caption{\label{fig:FTI_plot_M20} Violin plots for FTI testing parameters for a total mass of $20M_{\odot}$. The color scheme and the layout of the subplots are similar to Fig.~\ref{fig:TIGER_plot_M20}.}
\end{figure*}

\begin{figure*}
\includegraphics[width=0.99\linewidth]{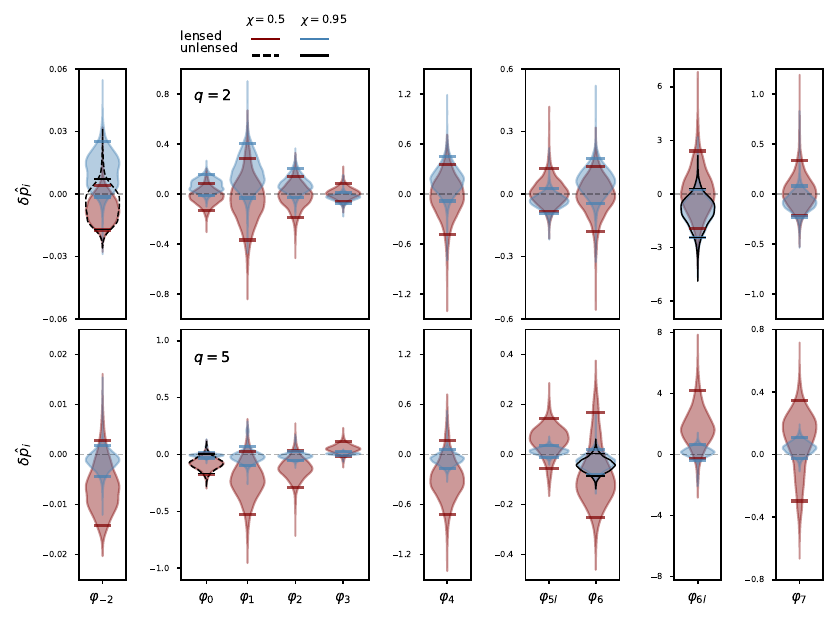}
	\caption{\label{fig:FTI_plot} Violin plots for FTI testing parameters for a total mass of $80M_{\odot}$. The color scheme and the layout of the subplots are similar to Fig.~\ref{fig:TIGER_plot_M20}.}
\end{figure*}

We give the posterior distributions of the FTI testing parameters for all our low and high mass injections in Figs.~\ref{fig:FTI_plot_M20} and~\ref{fig:FTI_plot} respectively. We also summarize the statistical level at which GR is excluded in Figs.~\ref{fig:quantile_plot_m20} and~\ref{fig:quantile_plot}.

\subsubsection{$20M_{\odot}$ injections}

Looking at Fig.~\ref{fig:FTI_plot_M20}, we observe significant deviations from GR when applying FTI to all the $M=20M_\odot$ binaries except the one with $q = 2$, $\chi = 0.5$, which is consistent with GR at the $90\%$ credible level for all testing parameters. While all cases with $q=2$ and $\chi=0.5$ are consistent with GR at the $90\%$ credible level, all the $q=2$, $\chi=0.95$ cases exclude GR (again at the $90\%$ credible level). GR is excluded at the $90\%$ credible level for all other binaries and testing parameters, except for the parameters $\varphi_{-2}$ and $\varphi_{7}$ for the $q=5$, $\chi=0.95$ binary.

In particular, the parameters $\varphi_{1}$ and $\varphi_{4}$ for $q=5$, $\chi=0.5$ show extreme deviations, with posteriors railing against the upper prior range of $60$ we used for the deviation parameters. We chose not to extend the prior range on these parameters further, as we find that all these large deviations are due to systematics due to the lack of precession in the current FTI implementation, rather than Type~II lensing. For these testing parameters, we also notice a railing against the prior bounds for the chirp mass and luminosity distance which we similarly do not correct by extending the ranges for the same reason.

To confirm that these deviations are indeed not due to lensing, we performed the FTI analysis on unlensed injections for the cases with the largest GR quantiles, as we did for TIGER, and also plot these in Fig.~\ref{fig:FTI_plot_M20}. The testing parameters chosen for the unlensed cases are $\varphi_{4}$, $\varphi_{6l}$ for $q = 2$ and $\varphi_{5l}$, $\varphi_{7}$ for $q = 5$, where the two cases for each mass ratio correspond to $\chi = 0.5$ and $0.95$, respectively. We find that the unlensed posteriors exhibit similar biases to their respective lensed counterparts, indicating that the observed biases are due to the absence of precession in the waveform currently used for the FTI test. Thus, we conclude that for the binaries considered here, precession can cause significant biases, but Type~II lensing does not lead to a significant additional bias beyond that due to precession. In order to further investigate the effects of precession, we applied the FTI test to an aligned-spin (\texttt{SEOBNRv4HM\_ROM}) injection for the binary with $q=2$ and $\chi=0.95$ and testing parameter $\varphi_{5l}$ since this gives us the highest GR quantile. We find that the posterior exactly peaks at the GR value of zero, thus verifying that the biases observed were indeed due to precession.

When we plot the FTI significances in Fig.~\ref{fig:quantile_plot_m20}, we find that most $q=5$ binaries exclude GR at such high credible levels that the estimation of the GR quantile becomes unreliable with the order of \(10^4\) posterior samples we obtain. Thus, as in~\cite{Narayan:2023vhm}, we conservatively report a lower bound of \(3\sigma\) on the significance of the deviations, since that study found that larger significances are not accurately obtained with this number of posterior samples.

The binary with $q=2$ and $\chi=0.5$ is consistent with GR at $<1.4\sigma$ across all testing parameters. However, all other binaries exclude GR at \(>2\sigma\), except for the parameters $\varphi_{-2}$ and $\varphi_{7}$ in the $q=5$, $\chi=0.95$ configuration, which are consistent with GR at well below \(1\sigma\). Notably, all testing parameters for the $q=5$, $\chi=0.5$ case, and all but the aforementioned two for the $q=5$, $\chi=0.95$, show deviations from GR exceeding \(3\sigma\). 

For the $q = 2$ injections, we recover smaller values for the primary spin magnitudes. More pronounced biases are observed in the $q = 5$ injections, particularly in the recovered mass ratios. Testing parameters $\varphi_{5l}$, $\varphi_{3}$, $\varphi_{2}$ recover more asymmetric mass ratios while the others recover more symmetric values, with an exception of $\varphi_{7}$ where the injected value is close to the median of the recovered posterior. Additionally, these higher mass ratio injections yield significantly larger chirp masses, total masses, and luminosity distances than the injected values. The FTI analysis of the $q=5$, $\chi=0.95$ injection also recovers lower values for the component spin magnitudes.

\subsubsection{$80M_{\odot}$ injections}
FTI finds that all lensed binaries with total mass $80M_{\odot}$ are consistent with GR at $90\%$ credibility for all testing parameters, though GR is excluded at up to $1.5\sigma$ in some cases. In particular, as illustrated in Fig.~\ref{fig:quantile_plot}, GR is excluded at $\gtrsim1\sigma$ for all testing parameters of the $q = 2$, $\chi=0.95$ binary and for all except $\varphi_{6}$ and $\varphi_{7}$ for the $q=5$, $\chi=0.5$ binary (which gives the $1.5\sigma$ exclusion for $\varphi_0$). The cases where GR is recovered in the tail of the distribution are likely due to the lack of precession in the current implementation of FTI, similar to the much larger biases found for most of the $M=20M_{\odot}$ injections. The lack of such significant bias in the $80M_{\odot}$ case is presumably due to the shorter signals in the sensitive band of the detectors making the effects of precession on the signal less important.

When applying FTI to binaries with $\chi=0.95$ for both mass ratios $q=2,5$, we observe that the posteriors peak at more symmetric mass ratios and smaller spin magnitudes, and the posteriors for the binary with $\chi=0.95$ and $q=5$ peak at larger luminosity distances. Additionally, for $q=2$, $\chi=0.95$, binary, we find a tendency for lower recovered chirp masses for testing parameters $\varphi_{-2}$, $\varphi_{0}$, $\varphi_{1}$, $\varphi_{2}$.

In order to study the effects of systematics, we analyze unlensed injections for the testing parameters that give the largest GR quantiles and make a comparison with the corresponding lensed cases, shown in Fig.~\ref{fig:FTI_plot}. Specifically, we consider \(\varphi_{-2}\), \(\varphi_{6l}\) for $q=2$ and \(\varphi_{0}\), \(\varphi_{6}\) for $q=5$, where the two values for each mass ratio correspond to \(\chi=0.5\) and \(0.95\), respectively. We observe that all the unlensed posteriors overlap significantly with the lensed posteriors, indicating that there are no significant additional biases due to Type~II lensing in these cases beyond the ones from precession.

\subsection{MDR}
\begin{figure*}
\includegraphics[width=0.99\linewidth]{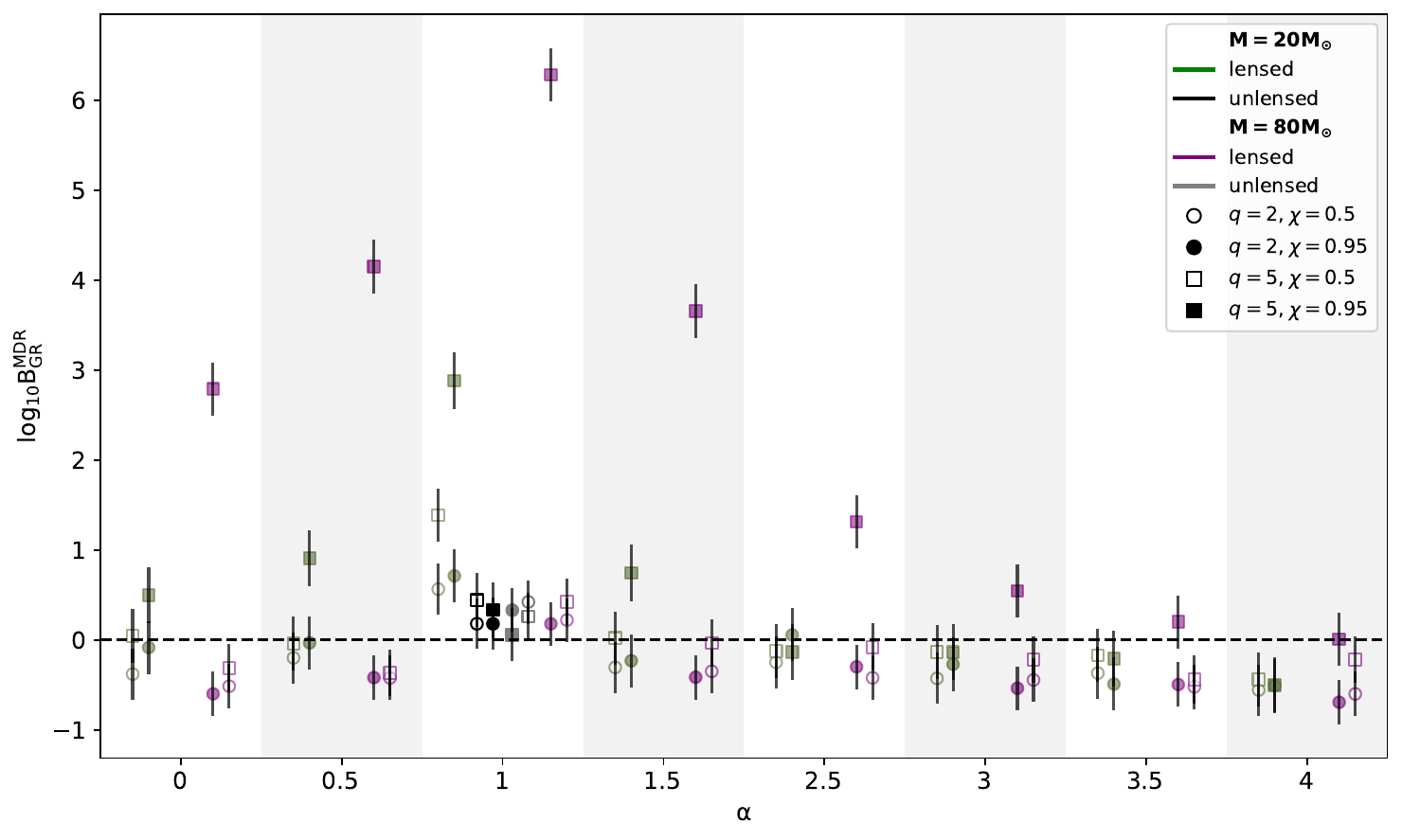}
\caption{\label{fig:MDR_plot} Plot of Bayes factors comparing MDR and GR for the set of eight discrete values of the modified dispersion relation exponent $\alpha$ we consider, excluding $\alpha = 2$ where there is no dispersion. The horizontal positioning of the points for each $\alpha$ value is just done for clarity.}
\end{figure*}

\begin{figure}
\includegraphics[width=\columnwidth]{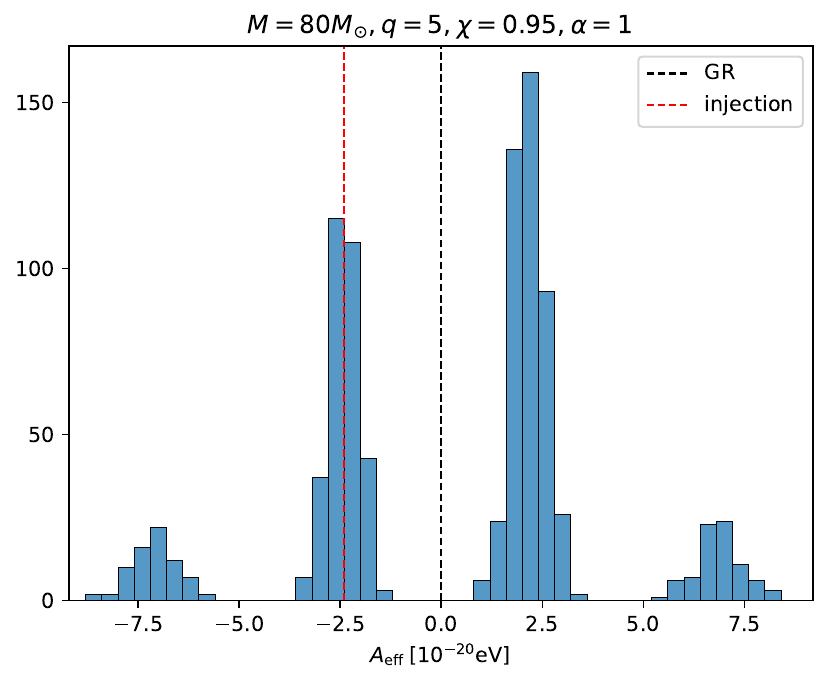}
	\caption{\label{fig:MDR_hist} $A_{\text{eff}}$ posterior for the $\alpha=1$, $M=80M_\odot$, $q = 5$, and $\chi = 0.95$ case. We also mark the GR value of zero and the injected value due to Type~II lensing phase shift of $\pi/2$. Here we have zoomed in around the GR value of zero where there is the largest posterior probability, but there are additional small peaks outside of this region, so only $17\%$ of the total probability is included in the region plotted.}
\end{figure}

For MDR, we report the results of the test in Fig.~\ref{fig:MDR_plot} as Bayes factors $\text{B}^\text{MDR}_\text{GR}$ comparing the MDR and GR models. The Bayes factor is the ratio of the evidences (i.e., the marginal likelihoods) of the two models. Larger values of $\log_{10}\text{B}^\text{MDR}_\text{GR}$ correspond to more support for MDR, while negative values correspond to support for GR. We quote Bayes factors instead of plotting the posteriors and quoting the GR quantiles because we find that the posteriors for $A_{\text{eff}}$ has multiple modes with no support at the GR value in the cases where GR is excluded at a high credible level. Thus, it would be misleading to quantify the statistical level at which GR is excluded from such a posterior using quantiles. An example of such a multimodal posterior is given in Fig.~\ref{fig:MDR_hist} for the binary with $M=80M_{\odot}$, $q=5$, and $\chi=0.95$ analyzed with $\alpha=1$.

The posterior of $A_{\text{eff}}$ is bimodal for all $\alpha$ values except $\alpha=1$, which gives the largest Bayes factor, and where there is an infinite $\pm \pi$ degeneracy around the Type II lensing phase shift value of $\pi/2$, as discussed in~\cite{Baka:2025drk}. Here infinite values of $A_{\text{eff}}$ reproduce the modified GW signal exactly through frequency independent phases $\pm(k+1/2)\pi$ for $k=(0,1,2,...)$. We observe this in Fig.~\ref{fig:MDR_hist} where the first mode of the $A_{\text{eff}}$ posterior at negative values, corresponding to a phase shift of $\pi/2$, peaks close to the injected value of $-2.4\times10^{-20}~\text{ eV}$. We also find that the two $\alpha$ values with the next largest support for MDR for a given binary, at least for the $q=5$, $\chi = 0.95$ cases, are $0.5$ and $1.5$, as would be expected, by continuity, and the support decreases as $\alpha$ becomes further from $1$.

As for TIGER and FTI, we also analyzed unlensed injections where the disagreement with GR is the largest. For all injections considered, this corresponds to $\alpha = 1$. We note that the values of $\log_{10}\text{B}^\text{MDR}_\text{GR}$ for all these cases are greater than zero, which is unexpected. To investigate this further, we used the Savage-Dickey ratio to estimate the Bayes factor by computing the ratio of the prior to the posterior at $A_{\text{eff}} = 0$ for the Bayes factor $\text{B}^\text{MDR}_\text{GR}$. We compared the $\log_{10}$ values of the Savage-Dickey ratio to the corresponding $\log_{10}\text{B}^\text{MDR}_\text{GR}$ and found that the $\log_{10}$ Savage-Dickey values were negative, indicating a potential discrepancy. For example, in the unlensed case with the highest $\log_{10}$ Bayes factor, corresponding to a system with total mass $20M_\odot$, mass ratio $5$, and spin $0.5$, the reported value is $0.4 \pm 0.3$, whereas the $\log_{10}$ of the Savage-Dickey ratio for this case is $- 0.1$. These findings suggest that \texttt{\textsc{Bilby}} may overestimate the evidence and/or underestimate its uncertainty.

%\textcolor{blue}{This potentially contributes to the values of the $\log_{10}$ Bayes factor of unlensed signals in Fig.~\ref{fig:MDR_plot} being $>0$ for $\alpha=1$.}  We also find that the two $\alpha$ values with the next largest support for MDR for a given binary, at least for the $q=5$, $\chi = 0.95$ cases, are $0.5$ and $1.5$, as would be expected, by continuity, and the support decreases as $\alpha$ becomes further from $1$. 

For the $80M_{\odot}$ injections, the high spin binaries recover biased individual spin magnitudes with both GR and MDR analysis recovering lower values than those injected. 

\subsection{IMR Consistency Test}

\begin{figure}
\includegraphics[width=\columnwidth]{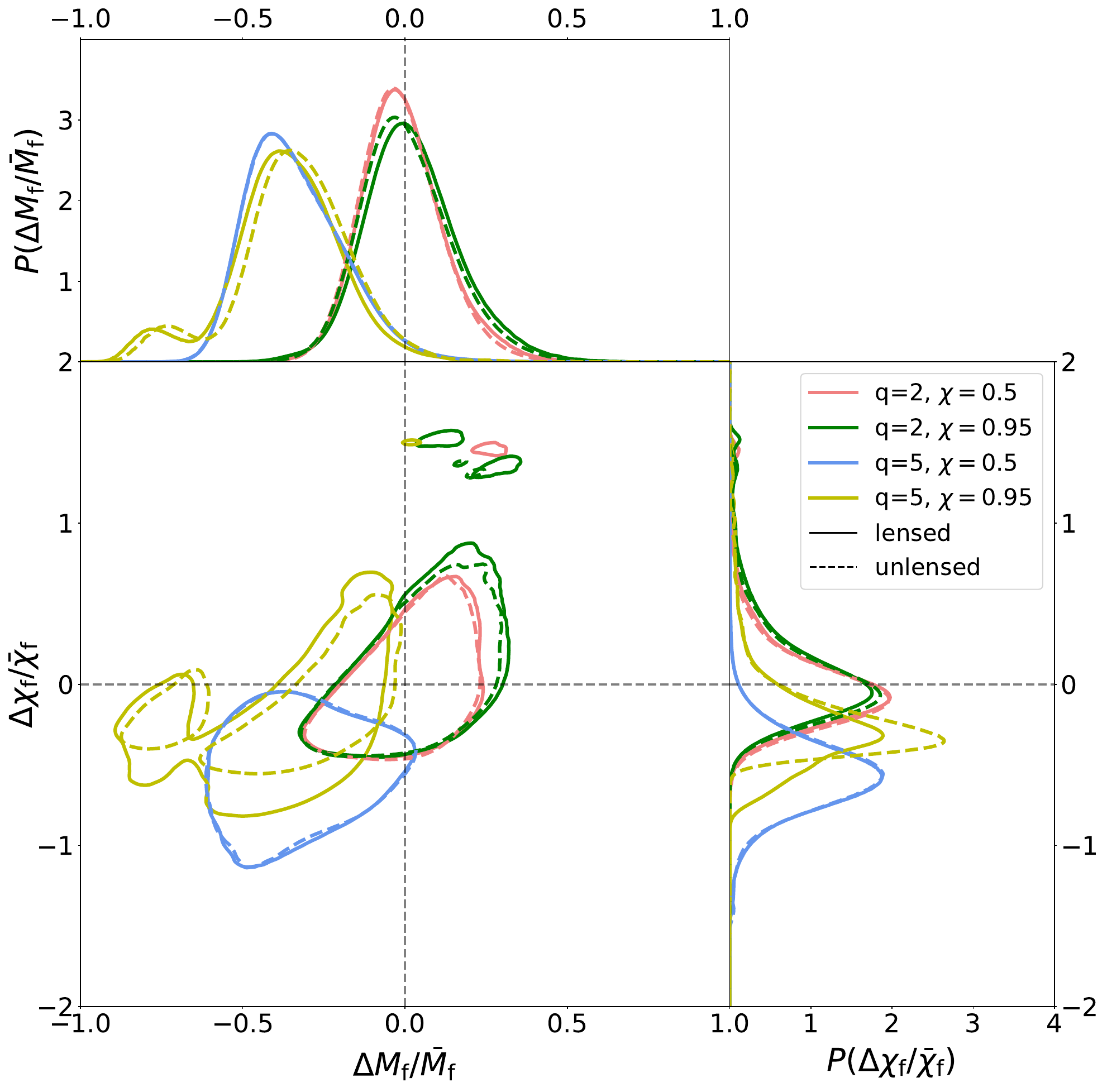}
	\caption{\label{fig:IMRCT_plot}The results from the IMR consistency test presented as the $90\%$ credible regions of the joint posterior distributions of the recovered final mass and spin deviation parameters for lensed and unlensed injections  for mass ratios $2$ and  $5$ with dimensionless spins of $0.5$ and $0.95$. We also show the one-dimensional distributions for the marginalized deviation parameters.}
\end{figure}

We perform the IMR consistency test only on the higher mass injection since the LVK only applies the test to signals with an SNR of at least $6$ in both the inspiral and postinspiral portions, since both need to be informative for the test to make sense (see~\cite{O2_TGR,O3a_TGR,O3b_TGR}). The cutoff frequencies demarcating the inspiral and postinspiral phases for $M=20M_{\odot}$ are $415$, $418$~Hz and $330$, $399$~Hz for the $q=2$ and $q=5$ binaries, respectively, where the two numbers for each mass ratio give the results for $\chi = 0.5$, $0.95$. With these cutoffs, the largest SNR in the postinspiral regime is $5.76$ for the $q=2$, $\chi = 0.95$ case, just below the threshold. For the $M=80M_{\odot}$ binaries, the cutoff frequencies are $104$, $108$~Hz and $80$, $110$~Hz for $q=2$ and $q=5$, respectively, where again the two values correspond to $\chi = 0.5$, $0.95$. The SNR in the post-inspiral regime range for the $80M_{\odot}$ binaries under consideration ranges from $\sim 10 \text{ to} \sim 15$.

In Fig.~\ref{fig:IMRCT_plot}, we show our results for the IMR consistency test on both lensed and unlensed injections. We find that the $q = 2$ lensed (unlensed) cases are consistent with GR [GR quantiles of $\sim17\% \,(\sim19\%)$, $\sim10\% \,(\sim13\%)$ for $\chi = 0.5$ and $0.95$, respectively]. The $q = 5$ cases show notable inconsistency (GR quantiles of $\sim93\%$, $\sim95\%$ for $\chi = 0.5$ and $0.95$), even without lensing (GR quantiles of $\sim94\%$, $\sim92\%$).

The waveform model used for injection and recovery in the IMR consistency test is \texttt{IMRPhenomXPHM} which includes higher order modes. Since we found a significant bias (GR quantiles of $>95\%$) in the IMR consistency test for (quasicircular) unequal-mass nonspinning binaries observed face-on in~\cite{Narayan:2023vhm}, we thus suspect that the bias observed for the $q = 5$ unlensed injections here could also be due to the presence of higher modes, even though we are considering an inclination angle of $\pi/3$, while ongoing studies~\cite{Mukesh_ICTS} have found that these biases are only significant for binaries very close to face-on (or face-off). We thus applied the IMR consistency test to the unlensed $q=5$, $\chi=0.95$ injection including only the $(2,\pm 2)$ modes in the coprecessing frame (i.e., using the \texttt{IMRPhenomXP} waveform model) for both the injection and recovery. We use the same cutoff frequencies as in the \texttt{IMRPhenomXPHM} analysis. We indeed find that the unlensed injection agrees with GR in this case (GR quantile of $\sim38\%$). Similar results were obtained in~\cite{Narayan:2023vhm} wherein GR deviations were found when a face-on quasi-circular NR injection was recovered using the quasi-circular \texttt{IMRPhenomXPHM} model with GR quantiles $>95\%$.

%%%%%%%%%%%%%%%%%
\section{Summary and Conclusions}
\label{sec:concl}
%%%%%%%%%%%%%%%%%

In thus study, we investigate the impact of neglecting Type~II strong gravitational lensing when performing some of the LVK's standard GR tests on lensed BBH signals. Specifically, we examine the response of the TIGER, FTI, MDR, and IMR consistency tests to simulated Type~II lensed BBH GW signals within the LIGO-Virgo network, assuming O5 design sensitivity. We simulate Type~II lensed signals for the precessing binaries under consideration. These binaries have redshifted total masses of $20M_{\odot}$ and $80M_{\odot}$, each with mass ratios of $2$ and $5$. For each mass ratio, we consider equal dimensionless spins (effective precession spin parameter, $\chi_{p}$) of $0.5$ ($0.45$) and $0.95$ ($0.85$), a total of eight cases, all with the same random spin angles that will lead to significant precession. Each binary is oriented to have an inclination angle of $\pi/3$ and placed at a luminosity distance so that it has a network SNR of $25$ with the magnification of $1$ that we consider. For each parameter combination, we analyze an unlensed BBH signal for comparison with the corresponding lensed case. This is done for a single testing parameter for TIGER/FTI and a specific $\alpha$ value for MDR. All the injections are made in zero noise.

For the TIGER test, all lensed binaries with \(M = 20 M_{\odot}\) were consistent with GR at the $90\%$ credible level. However, for higher mass binaries (\(M = 80 M_{\odot}\)), significant deviations were found for the configuration with \(q = 5\) and \(\chi = 0.95\), where GR was excluded at \(> 2\sigma\) for some testing parameters. For the FTI test, very significant deviations from GR were found for binaries with \(M = 20 M_{\odot}\) and \(q = 5\). However, these deviations are also found for the unlensed injections and are thus due to systematics from analyzing strongly precessing injections with the current FTI analysis that does not include precession. These findings are consistent with the results in~\cite{Chandramouli:2024vhw}, who report similar biases in parametrized tests of GR due to unmodeled precession. Specifically, they consider a \(20 M_{\odot}\) binary with comparable SNR (\(30\)) and \(\chi_{p}\) values (\(0.45\) and \(0.9\)), as well as a mass ratio of \(1.5\). For this system, they find similar deviations to those we observe for the $20 M_{\odot}$ binary with a mass ratio of \(2\) when analyzing inclination angles of \(\pi/4\) and \(\pi/2\), which bracket our \(\pi/3\) inclination. All binaries with \(M = 80 M_{\odot}\) are consistent with GR at $90\%$ credibility for all FTI testing parameters.

For the MDR test, we find Bayes factors favoring the MDR model over GR for the \(q = 5\), $\chi = 0.95$ binaries, with MDR being favored more strongly for the \(M = 80 M_{\odot}\) binary. The strongest support for MDR was for \(\alpha = 1\), as expected, since this is where the MDR phase shift is able to reproduce the Type~II lensing phase shift exactly. However, GR is favored slightly for some of the $q=2$ and/or $\chi = 0.5$ cases.

Finally, for the IMR consistency test, we only analyze the binaries with $M = 80 M_{\odot}$, since the $20M_\odot$ binaries do not have sufficiently high postinspiral SNR. Here we observe that the $q = 2$ cases (with spins $0.5$ and $0.95$) are consistent with GR. However, the $q = 5$ cases exhibit significant inconsistencies, with GR quantiles as large as $95\%$. Furthermore, we also find these biases for the unlensed $q = 5$ injections, and find that they can be attributed to the influence of higher-order modes, as the quantiles reduce to $\sim40\%$ when using a waveform model limited to the $(2,\pm 2)$ modes in the coprecessing frame for both injection and recovery. These biases are likely related to similar biases found in~\cite{Narayan:2023vhm} when including higher-order modes in the analysis of face-on (quasicircular) nonspinning unequal-mass binaries.

The results of this paper highlight the necessity of ruling out strongly lensed binaries as a source of false GR violations before claiming a potential deviation from GR. Specifically, it is important to consider the potential for a Type~II lensed signal to mimic a false deviation from GR, especially in cases where the LVK lensing analyses (carried out simultaneously with the tests of GR) fail to identify Type~II signals or their companion lensed images. Addressing the risk of misinterpreting a Type~II lensed signal as a GR deviation requires incorporating the Type~II lensing effect into the baseline GR waveforms used in GR tests. A natural extension of this study is to evaluate the response of Type~II lensing analyses to the signals considered here and confirm that including the lensing effect in baseline GR waveforms resolves the biases.

%%%%%%%%%%%%%%%%%

\acknowledgments

We thank Anuj Mishra, Tomek Baka, and Rico Lo for useful comments and all the LIGO-Virgo-KAGRA testing GR group members who implemented the tests of GR we consider in publicly available code. PN is supported by NSF grant PHY-2308887; NKJ-M is supported by NSF grant AST-2205920; and AG is supported in part by NSF grants PHY-2308887 and AST-2205920. The authors are grateful for computational resources provided by the LIGO Laboratory and supported by National Science Foundation Grants PHY-0757058, PHY-0823459, PHY-1700765, and PHY-1626190. We also acknowledge the use of the Maple cluster at the University of Mississippi (funded by NSF Grant CHE-1338056) during the initial phase of this work.

This study used the software packages \texttt{\textsc{Asimov}}~\cite{Williams:2022pgn}, \texttt{\textsc{Bilby}}~\cite{Ashton:2018jfp}, \texttt{\textsc{bilby\_pipe}}~\cite{Romero-Shaw:2020owr}, \texttt{\textsc{bilby\_tgr}}~\cite{2024zndo..10940210A}, \texttt{\textsc{hanabi}}~\cite{2023PhRvD.107l3015L}, \texttt{\textsc{Dynesty}}~\cite{Speagle_2020}, \texttt{\textsc{LALSuite}}~\cite{LALSuite}, \texttt{\textsc{Matplotlib}}~\cite{Hunter:2007ouj}, \texttt{\textsc{NumPy}}~\cite{Harris:2020xlr}, \texttt{\textsc{PESummary}}~\cite{Hoy:2020vys}, \texttt{\textsc{Positive}}~\cite{London:2018nxs}, \texttt{\textsc{SciPy}}~\cite{Virtanen:2019joe}, and \texttt{\textsc{Seaborn}}~\cite{Waskom:2021psk}.

This is LIGO document number LIGO-P2400578.
\bibliography{lensing_tgr_refs}
\end{document}